# Quantification of flexibility from the thermal mass of residential buildings in England and Wales

**Dr Alexandre Canet** - Cardiff University, Cardiff, CF24 3AA, UK
**Prof. Meysam Qadrdan** - Cardiff University, Cardiff, CF24 3AA, UK

## Abstract
The increased integration of variable renewable generation into the power systems, along with the phase-out of fossil-based power stations, necessitate procuring more flexibility from the demand sectors. The electrification of the residential heat sector is an option to decarbonise the heat sector in the United Kingdom. The inherent flexibility that is available in the residential heat sector, in the form of the thermal inertia of buildings, is expected to play an important role in supporting the critical task of short-term balancing of electricity supply and demand. This paper proposes a method for characterising the locally aggregated flexibility envelope from the electrified residential heat sector, considering the most influential factors including outdoor and indoor temperature, thermal mass and heat loss of dwellings. Applying the method to England and Wales as a case study, demonstrated a significant potential for a temporary reduction of electricity demand for heating even during cold weather. Total electricity demand reductions of approximately 25 GW to 85 GW were shown to be achievable for the outdoor temperature of 10°C and -5°C, respectively. Improving the energy performance of the housing stock in England and Wales was shown to reduce the magnitude of available flexibility to approximately 18 GW to 60 GW for the outdoor temperature of 10°C and -5°C, respectively. This is due to the use of smaller size heat pumps in the more efficient housing stock. However, the impact of the buildings' retrofit on their thermal mass and consequently on the duration of the flexibility provision is uncertain.

## 1   Introduction

In the United Kingdom (UK), the decarbonisation of the economy is planned to be supported by the uptake of low carbon electricity generations and the electrification of services such as heating and transport. According to the Great Britain's electricity system operator, National Grid [1], by 2050, the electricity demand is expected to increase between 50% to 100%, whilst between 78% to 87% of the total electricity will be supplied by variable sources such as wind turbines and photovoltaics panels.

To compensate for the variable electricity generation and to address the challenge of balancing electricity supply and demand, the magnitude of flexibility required by the future power system will increase [2,3]. In the context of the power system operation, the term *flexibility* refers to the ability of the system to always balance electricity supply and demand in response to any changes in the expected generation and consumption. The continuous balancing of supply and demand can be achieved by modifying electricity production and/or consumption. In 2020, 45 GW of the 60 GW of flexibility required came from thermal generating units [1]. However, significantly higher flexibility required by the power system in 2050 (approximately two to four times higher [1]) is expected to be procured from low carbon means such as low carbon controllable generation, electrolysis, electricity storage, vehicle to grid and demand side response (DSR). The UK electricity system operator also recognises the potential for shifting the demand by using smarter controls for electricity based heating systems in combination with thermal storage [4].

Several studies investigated the provision of flexibility from the thermal mass of dwellings using heat pumps. A review of power-to-heat options to integrate renewable energy identified heat pumps and using thermal mass of buildings as the most favourable options[5]. A framework identified the energy generation capability of a building and its thermal mass as two major contributors to account for when quantifying DSR for residential and commercial buildings[6]. Authors in [7] used the EnergyPLAN software and showed the thermal mass of buildings is the most cost effective storage system when using heat pumps. Another study, in which the authors used OpenIDEAS/Modelica for their analysis [8], demonstrated that for a single-family dwelling equipped with an Air Source Heat Pump (ASHP), the use of the thermal mass of the dwelling decreases the electricity consumption of the ASHP by 75% to 94% during peak times [9]. The impacts on the level of comfort of providing flexibility with heat pumps in dwellings without thermal storage tanks was studied for two houses with different level of thermal mass using the thermal-dynamic simulation software TRNSYS [10]. It was shown that the house with higher thermal mass provided more thermal comfort and higher available power for flexibility services. The relation between thermal mass and amount of flexibility was also demonstrated for residential building located in a cold climate [11].

For the state of California, the aggregated magnitude of demand response that could be provided by heat pumps over a 15 minutes duration was estimated to be 9 GW, whilst the peak power capacity was ~40 GW in 2014 [12]. The impacts of insulation on the magnitude of flexibility available were investigated by comparing poorly insulated and well-insulated buildings in Denmark. The results showed that the heat load modulation could be large for poorly insulated buildings but only done for a short period of time (2 to 5 hours) whereas for well-insulated buildings the magnitude of flexibility will be low but could be provided for a longer timeframe. The provision of frequency response services using domestic heat pumps was explored in the literature [13,14].

A virtual thermal storage approach was used to represent the building stock and study the potential of heat pumps for demand side management in Germany [15]. However, this representation of the building stock does not seem to capture the non-homogeneity of the building stock such as differences in thermal losses and thermal mass between buildings, and thus overestimates the potential for flexibility provision. A generic quantification method was implemented using OpenIDEAS/Modelica to estimate the storage capacity of the thermal mass of buildings without compromising the comfort of the occupants [16]. This study showed that the insulation level, the type of heating systems and the duration of the demand side response event were key parameters when providing flexibility to the grid through the modulation of the heating system.

The majority of the literature looks at the level of flexibility that can be provided by single buildings or a small group of buildings or focuses on flexibility services with specific duration. There is a wide range of modelling techniques for quantifying the potential of inherent flexibility of buildings used in the literature including energy hub models [17], commercial and custom transient models but no specific technique appears to be prevalent. In this paper, we aim to use a transient model to characterise the magnitude and duration of technically available flexibility that can be provided by the thermal mass of the residential buildings using Air Source Heat Pumps (ASHPs) in England and Wales. The rationale behind focusing on ASHPs in this paper was based on the decarbonisation pathways for Great Britain (GB) published by the GB electricity system operator [4]. In the three pathways leading to net-zero by 2050, ASHPs are a dominant technology in comparison to ground source heat pumps (GSHPs) and direct electric heaters. Furthermore, our analysis attempts to investigate the scale of flexibility available from thermal inertia of buildings which is an under-studied research area, for this reason we excluded hot water tank and other thermal energy storage technologies. Finally, detailed input data and outputs from our analysis are published and available online, which will allow other researchers and users to produce new sets of results for different scenarios such as more efficient ASHPs, combination of ASHPs/GSHPs or only direct electric heaters.

Key contributions of this paper are:

- a methodology was developed to characterise the thermal parameters of the dwelling stock from Energy Performance Certificates,
- a methodology was developed to quantify the magnitude and duration of flexibility from the residential heat sector for local areas (known as Lower Layer Super Output Areas: LSOAs) across England and Wales.
- an investigation of the impacts of buildings retrofits on the amount of flexibility available from the dwelling stock was carried out.

## 2   Methods

Figure 1 shows the methodology used for this study. The first step was to create a database of thermal characteristics of dwellings for each LSOA in England and Wales using Energy Performance Certificates collected from the open data communities platform [18] and the information in the Standard Assessment Procedure 2012 [19] guidelines which are used for building regulation compliance in the UK. Using the thermal characteristics of dwellings in a thermal model of buildings, the magnitude and duration of two DSR flexibility services were characterised for a case study where all the dwellings in England and Wales have ASHPs installed:

1. Positive flexibility – an increase in the electricity consumption of heat pumps when all heat pumps increase their outputs to their maximum capacity. This provides a demand increase service to the public electricity network.
2. Negative flexibility – a decrease in the electricity consumption of heat pumps when all heat pumps are switched off. This provides a demand reduction service to the public electricity network.

The above naming convention was chosen for the sake of simplicity. Different names for such services might be used in the flexibility market.

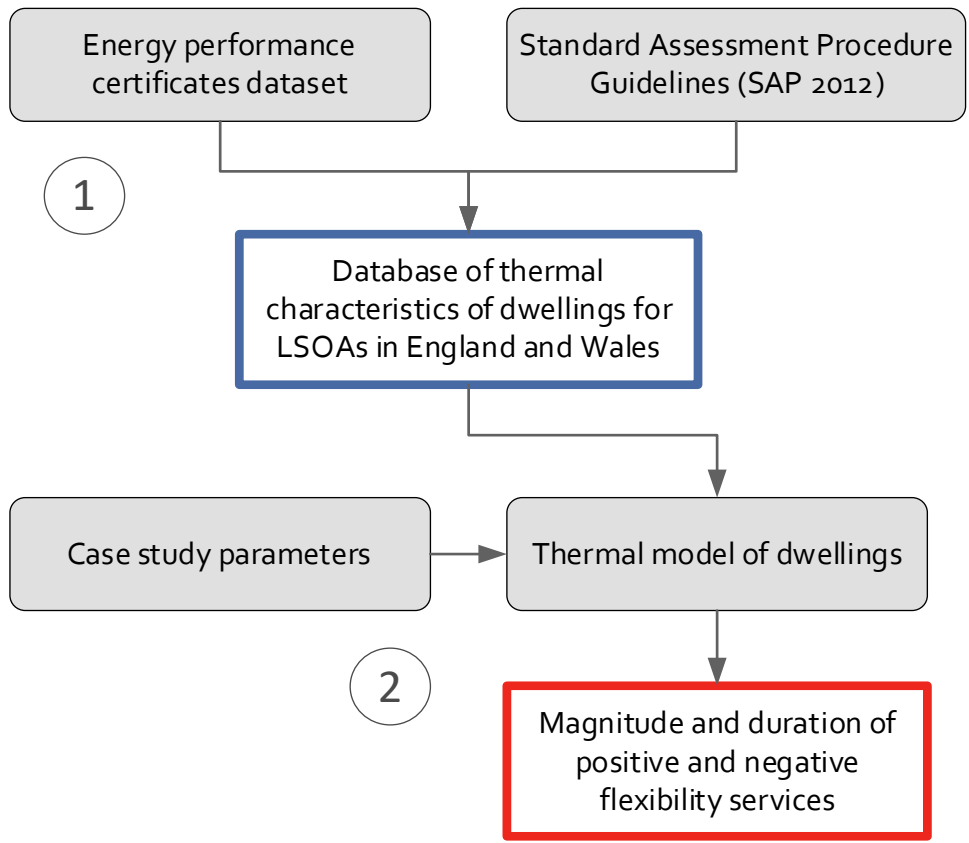

*Figure 1: Overview of the methodology.*

The methods describe in the following were implemented in Python [20]. The code is available (see details in section 6).

The dwelling stock of England and Wales used in this study is distributed over 34,753 LSOAs and 16 dwelling categories. A *dwelling category* is the combination of a dwelling type (i.e., detached, semi-detached, terraced, or flat) and a heating system (i.e., natural gas boiler, resistance heater, biomass boiler or oil boiler).

Figure 2 shows an overview of the steps used to calculate the magnitude and duration of the flexibility services for each dwelling category. After the indoor and outdoor air temperatures are set, they were used to calculate the current heating output of the dwelling to maintain the indoor air temperature constant. The current heating output of the ASHP at the set outdoor air temperature were used to derive the magnitude of the positive and negative flexibility services. In the last step, the duration for which the flexibility services can be provided are calculated.

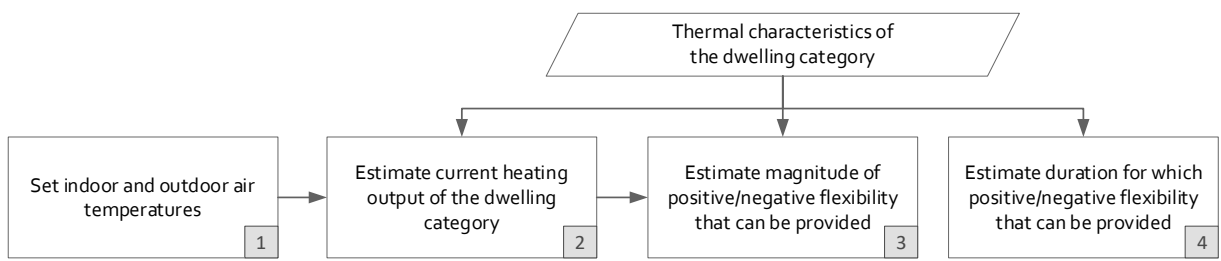

*Figure 2: Overview of the method to calculate the magnitude and duration of the flexibility services that can be provided to the main electricity network.*

## 2.1 Thermal characteristics of dwellings

For each LSOA, the average thermal losses, average size of the heating systems and the average thermal capacity were calculated for the 16 dwelling categories defined previously.

### 2.1.1 Calculating the thermal losses of dwellings and sizing air source heat pumps

The thermal losses or each dwelling category in each LSOA were derived using Equation 1.

$$QL_{d,l} = \frac{AQ_{d,l}}{HDH} \tag{1}$$

Where, $QL\ [kW/°C]$ is the thermal losses, $d$ is the dwelling category, $l$ is the target LSOA, $AQ\ [kWh]$ is the average annual heat demand and $HDH\ [°C * hours]$ is the number of heating degree hours in the region of the LSOA $l$ (see Table 1).

### 2.1.2 Calculating the thermal capacity of dwellings

The thermal capacity was calculated using Equation 2.

$$C_{d,l} = TFA_{d,l} \times SC \tag{2}$$

Where, $C\ [kJ/K]$ is the thermal capacity, $TFA_d\ [m^2]$ is the average floor area of the dwelling category $d$ in the LSOA $l$ and $SC\ [kJ/m^2/K]$ is the specific thermal capacity value.

### 2.1.3 Input data

An input dataset was created to calculate the thermal losses and thermal capacity of the dwelling stock. It includes for each LSOA and each dwelling category:

- the average annual heat demand before energy efficiency measures,
- the average annual heat demand after energy efficiency measures,
- the average total floor area,
- the specific thermal capacity,

The average annual heat demand before and after energy efficiency measures for each dwelling category in each LSOA in England and Wales published on the UKERC Energy Data Centre was used [21]. Using the same approach, average floor area for each building archetype in each LSOA was calculated using data available in Energy Performance Certificate of buildings.

The outliers for the average annual heat demand and floor areas for each dwelling category and each LSOA were dealt by:

- Replacing values where heat demand/floor area values is above the 99th percentile such as:
    - Heat demand value is the heat demand value of the 99th percentile.
    - Floor area value is the floor area value of the 99th percentile.
- Replacing values where heat demand/floor area values below the 1st percentile such as:
    - Heat demand value is the heat demand value of the 1st percentile.
    - Floor area value is the floor area value of the 1st percentile.

The thermal capacity level called "medium" with a specific thermal capacity of 250 $kJ/m^2/K$ published in SAP 2012 [19] was used in this study.

Table 1 shows the number of heating degree days and the design temperature of heating systems in each region of England and Wales. A lookup table linking each LSOA to a region was used to assign the number of heating degree days and design temperature.

*Table 1: Number of heating degree days and design temperature of heating systems in regions of England and Wales. The heating degree days were calculated based on a 15.5°C temperature and the monthly average temperature from SAP 2012[19]. The design temperatures of heating systems were derived from the microgeneration installation standard 3005 [22].*

| Region | Heating degree days | Design temperature (°C) |
| --- | --- | --- |
| East | 1,873.6 | -3 |
| East Midlands | 2,055.7 | -3 |
| London | 1,773.5 | -2 |
| North East | 2,216.8 | -5 |
| North West | 2,359.8 | -5 |
| South East | 1,815.7 | -1 |
| South West | 1,740.6 | -2 |
| Wales | 2,058.9 | -3 |
| West Midlands | 2,055.7 | -3 |
| Yorkshire and The Humber | 2,216.8 | -5 |

## 2.2 Sizing of air source heat pumps

The sizing of the ASHPs in the dwelling stock was derived using Equation 3.

$$S_{d,l} = \Delta T \times QL_{d,l} \qquad (3)$$

Where, $S\ [kW]$ is the size of the ASHPs and $\Delta T$ is the temperature difference between the indoor design temperature, which was set at 21°C, and the outdoor design temperature of the heating system (see Table 1).

## 2.3 Estimating the magnitude and duration of flexibility services

A lumped parameter model (1R 1C) [23] was used to create a thermal model of a dwelling which was used to:

1. Calculate the magnitude of the flexibility service that can be provided, and,
2. Calculate the duration for which the flexibility service can be provided.

Figure 3 shows a diagram of the RC model used. Equations 4 describe the heat balance equation of this model[23].

$$P(t) - \frac{1}{R_{th}}(T_{indoor}(t) - T_{outdoor}) = C_{th}\frac{dT_{indoor}(t)}{dt} \qquad (4)$$

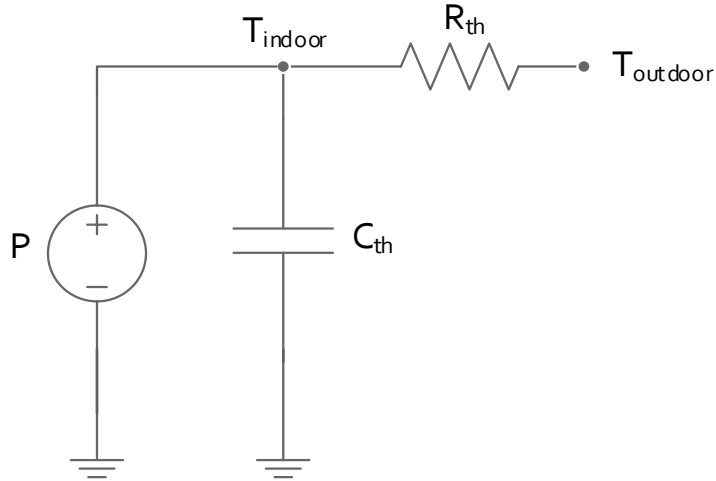

*Figure 3: Resistor-capacitor (RC) model with a single R and a single C used as a thermal building model of a dwelling. $R_{th}$ [°C/W] is the thermal resistance, and is the inverse of the thermal losses of the dwelling, $C_{th}$ is the thermal capacitance [J/°C] of the dwelling, $T_{indoor}$ is the indoor air temperature [°C], $T_{outdoor}$ is the outdoor air temperature [°C] and P is the heating output [W] from the ASHP.*

### 2.3.1 Magnitude of the flexibility service

The magnitude of positive and negative flexibility that can be provided were calculated using Equation 5 and 6.

$$PF = \frac{MQ - IQ}{COP_{T_{outdoor}}} \quad (5)$$

$$NF = \frac{-IQ}{COP_{T_{outdoor}}} \quad (6)$$

Where, $IQ$ [W] is the initial heating output, $MQ$ [W] is the maximal heating output of the heating system, $PF$ [W] is the magnitude of positive flexibility, $NF$ [W] is the magnitude of negative flexibility and $COP_{T_{outdoor}}$ is the coefficient of performance of the ASHPs at the outdoor air temperature $T_{outdoor}$.

The initial heating output is the output of the heating system to maintain the indoor air temperature constant. It is defined by equation 7.

$$IQ = \frac{T_{indoor} - T_{outdoor}}{R_{th}} \quad (7)$$

Table 2 shows the average COP of ASHPs for the outdoor air temperatures used in this study.

*Table 2: Outdoor air temperature and average coefficient of performance of ASHPs used in this study.*

| Outdoor air temperature in England and Wales | Average COP used in this study based on ASHPs |
|---|---|
| -5°C | 2[24] |
| 0°C | 2.3[25] |
| +5°C | 2.4[25] |
| +10°C | 2.6[25] |

### 2.3.2 Duration of the flexibility service

To calculate the duration of the flexibility service, it was determined how long the heating output $P$ [W] can be kept at a given value before the higher or lower limits of indoor temperature $T_{indoor\ limit}$ [°C] are reached. The outdoor air temperature $T_{outdoor}$ [°C] was assumed to stay constant.

The lower and higher limits for $T_{indoor\ limit}$ were based on literature data and summarised in Equation 8. The low indoor air temperature threshold was fixed at +18°C following recommendations from Public Health England[26]. The high indoor air temperature threshold was fixed at +24°C as temperatures above could cause discomfort and potential harms[27].

$$18°C \leq T_{indoor} \leq 24°C \tag{8}$$

The heat balance Equation 4 was used to derive Equation 9 which is used to calculate the duration $D$ of a flexibility service based on the parameters of the thermal building model and the heating output of the ASHP.

$$D = \frac{\ln(T_{indoor\ limit} - \frac{A}{1-B}) - \ln(T_{indoor} - \frac{A}{(1-B)})}{\ln B} \tag{9}$$

Where, $A$ is shown in Equation 10 and $B$ is shown in Equation 11.

$$A = \frac{T_{outdoor}}{R_{th}C_{th}} + \frac{P}{C_{th}} \tag{10}$$

$$B = 1 - \frac{1}{R_{th}C_{th}} \tag{11}$$

Figure 4 shows the process followed to calculate the duration of a positive flexibility service for a dwelling for specific case when Equation 9 is not valid which includes:

- When the initial indoor temperature $T_{indoor}$ of the dwelling is above the maximum limits of the indoor air temperature, the duration of the service will be 0s, and,
- When based on the magnitude of the positive flexibility service provided, the minimum or maximum indoor air temperature $T_{indoor\ limit}$ will never be reached, the duration of the service will be infinite.

A similar process is conducted when providing a negative flexibility service.

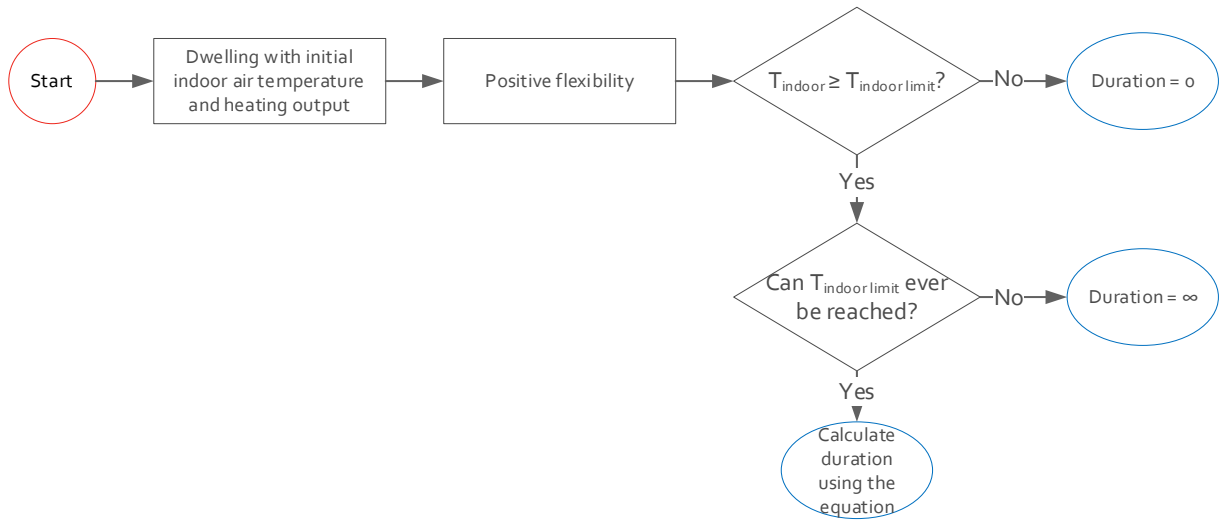

*Figure 4: Process to calculate the duration of a positive flexibility service for a dwelling.*

## 3 Results

The methods described in this paper were demonstrated on the dwelling stock of England and Wales in 2018 which comprises 23.4 million dwellings with a total heat demand of 350 TWh per annum[21]. Figure 5 shows the number of dwellings by dwelling forms and heating systems in the dwelling stock. More than 85% of the dwellings have gas boilers, 9% resistance heating and the rest oil and biomass boilers. Semi-detached houses represent 31% of the dwellings and the rest is distributed almost equally into detached houses, terraced houses and flats. The flexibility from dwellings was calculated for a scenario where the heating systems in 100% of the dwellings in the dwelling stock were converted to ASHPs.

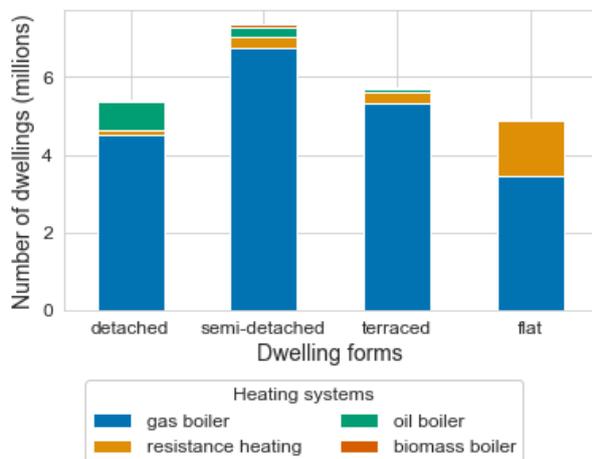

*Figure 5: Number of dwellings in England and Wales in 2018[21].*

### 3.1 Thermal characteristics of dwellings in England and Wales

For each dwelling category in each LSOAs, the thermal losses and the thermal capacity were calculated using methods described in Sections 2.1.1 and 0. Due to lack of accurate data, three levels of thermal capacity were calculated named: low, medium and high. In the rest of this paper, if not stated otherwise, the medium thermal capacity values were used to produce the results.

Figure 6 shows the distribution of the average thermal capacity and thermal losses of dwellings per dwelling form. Flats have on average, lower thermal capacity and lower thermal losses. Detached houses have higher thermal losses and higher thermal capacity. Semi-detached and terraced houses have similar characteristics.

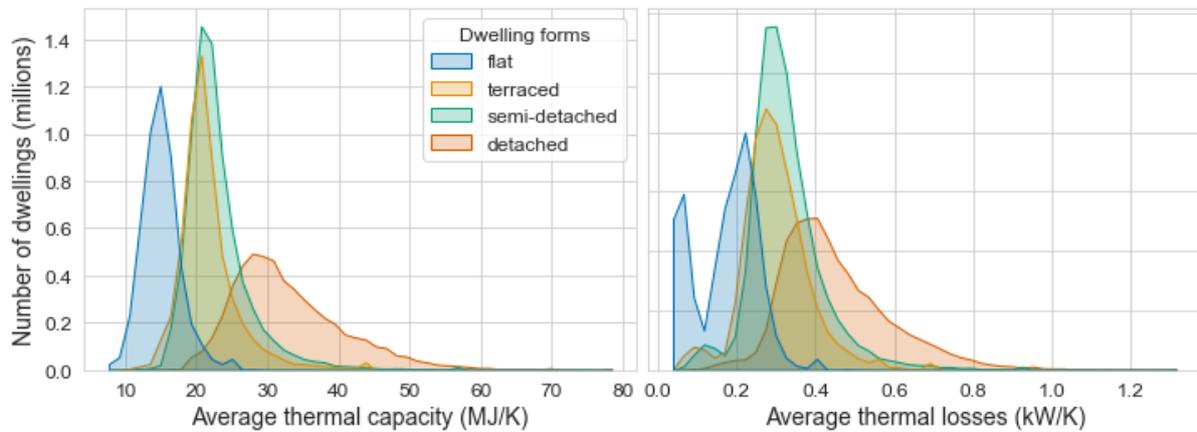

*Figure 6: Distribution of the thermal characteristics of four dwelling forms in England and Wales. The average thermal capacity is based on medium thermal capacity level. The figures were smoothed for visualization purposes by grouping the values into 50 bins.*

The estimated thermal losses of the dwellings, the indoor and outdoor design temperature in each geographical region were used to calculate the installed capacity of the ASHPs (see Section 2.2). The use of backup heating solutions and/or thermal storages could impact the sizing of ASHPs but were not considered in this study. Figure 7 shows the total capacity of ASHPs for different dwellings forms, assuming all buildings will install ASHPs. In total, it was estimated that the total capacity of ASHPs that will be installed is 176 GW$_{thermal}$.

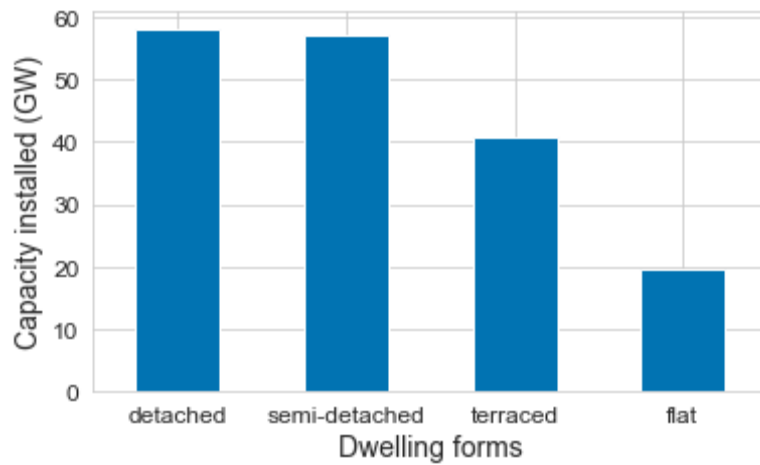

*Figure 7: Capacity installed of residential heating systems in England and Wales for different dwelling forms.*

## 3.2  Flexibility from dwellings in England and Wales

The magnitude and duration of providing flexibility (i.e. adjusting the electricity consumption of heat pumps) to the power system were calculated for the England and Wales dwelling stock assuming that all dwellings have ASHPs.

Figure 8 shows the magnitude and duration of the positive and negative flexibility services for four outdoor air temperatures -5, 0, +5 and +10°C, considering that all dwellings had the same initial indoor air temperature of +19°C.

The orange line with circle marker shows the results for an outdoor air temperature of 0°C. The positive flexibility can be provided for "unlimited" duration as even if we are increasing the outputs of the heat pumps to their maximum, the maximum indoor temperature of +24°C will never be reached. This is because the size of the heat pump was selected to compensate for heat losses of the buildings for a temperature gradient of almost +24°C. Depending on the region, the outdoor design temperature of the heating systems varies from -1°C to -5°C (see Section Methods). The demand decrease can be provided for less than an hour before the indoor air temperature of the dwellings reaches the minimum indoor temperature of +18°C.

At outdoor temperature of -5°C, the heating systems in all the dwellings are working at almost maximum capacity to maintain the indoor air temperature of +19°C. Hence, close to 100% (ca. 87 GW – considering COP of 2) of the capacity installed is available to provide negative flexibility.

The magnitude of positive flexibility increases with the outdoor air temperature but the duration for which it can be provided decreases. It is explained because:

- At higher outdoor temperature, heat pumps operate at reduced capacity to maintain the desired indoor temperature. This means larger spare capacity is available to ramp up.
- At higher outdoor temperature, running the heating systems at maximum capacity makes the indoor temperature to reach the maximum set limit faster.

The opposite is observed with the magnitude of negative flexibility that can be provided and its duration and how it varies with outdoor air temperature.

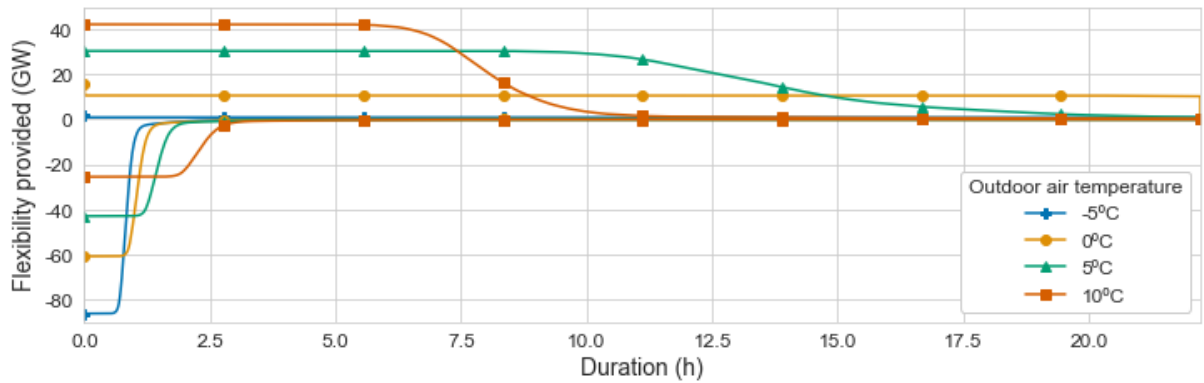

*Figure 8: Estimated magnitude and duration of flexibility services provided when the initial indoor air temperature in dwellings is +19°C.*

Figure 9 shows a map of the aggregate magnitude of positive and negative flexibility services for the outdoor air temperature of +5°C by local authority. The local authority with the darker colours represents the areas with the higher capacity of heating systems installed. The number of dwellings and their characteristics are the main factors to explain the differences between local authorities.

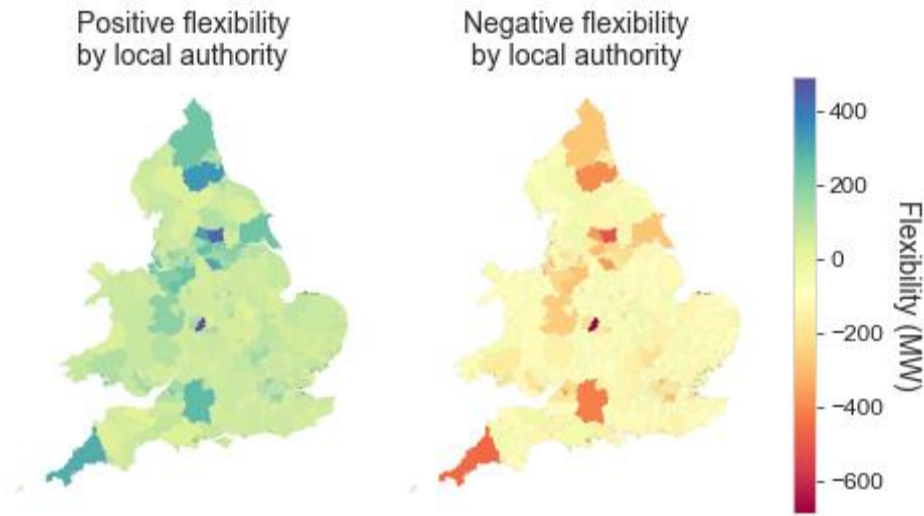

Figure 9: Maps of the positive and negative flexibility at local authority level of England and Wales. These maps are based on an indoor air temperature for all dwellings of +19°C and an outdoor air temperature of +5°C. The distribution of the dwellings in England and Wales was extracted from the dwelling stock dataset[21].

## 3.3 Parametric sensitivity analysis

In the following, the sensitivity of the results to the choice of the indoor air temperature of the dwellings and the impact of thermal losses and the thermal capacity were assessed.

Figure 10 shows the results when setting the indoor air temperature of all dwelling to +20°C. The magnitude of positive flexibility is decreased, and the magnitude of negative flexibility is increased compared to when indoor air temperature was set at +19°C (Figure 9 ).

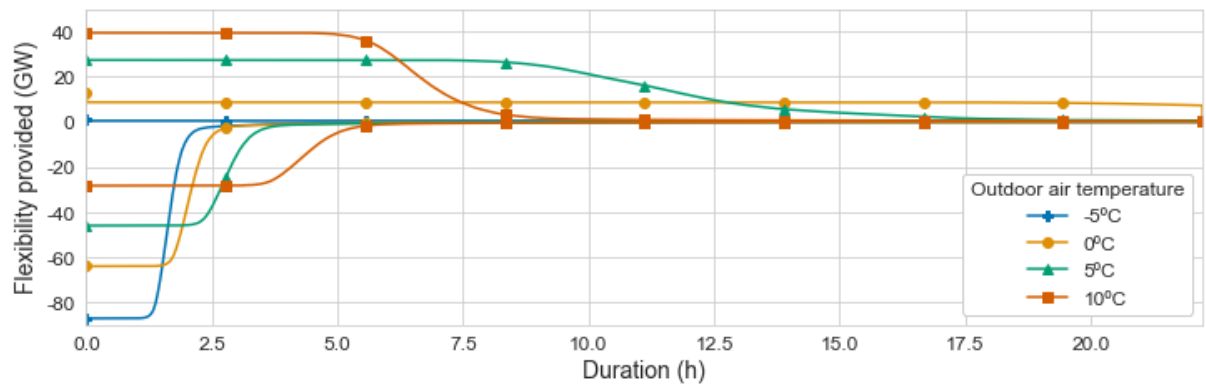

Figure 10: Estimated magnitude and duration of flexibility services provided when the initial indoor air temperature in dwellings is +20°C.

### 3.3.1 Impact of accounting for diversity of the indoor air temperature of dwellings

The previous results represent scenarios where all dwellings have the same indoor air temperature, however it is unlikely to happen. To estimate the impact of having a different indoor air temperature in each dwelling in England and Wales on the magnitude of flexibility that can be provided, the probability density function (PDF) shown in Figure 11 was used to assign an indoor air temperature to each dwelling.

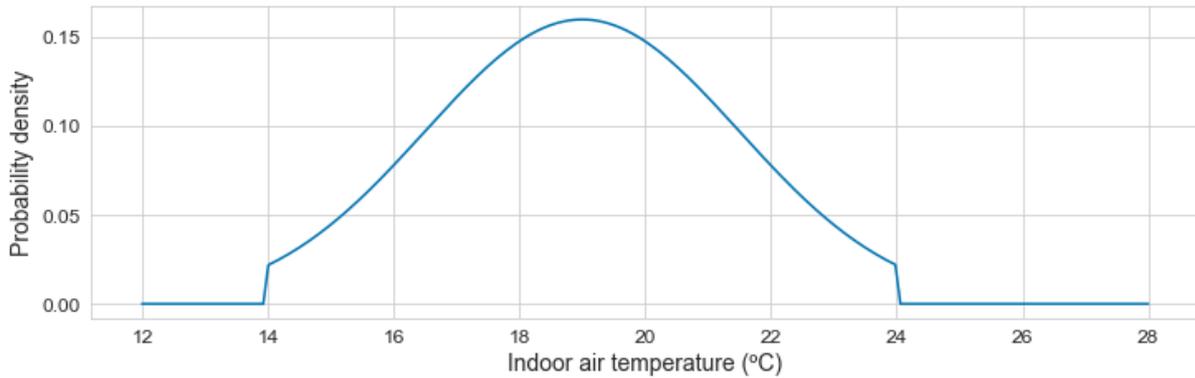

*Figure 11: Probability density function used to assign an indoor air temperature to every dwelling in England and Wales. Truncated normal distribution at +14 and +24˚C with a mean value of +19˚C and a standard deviation of 2.5˚C. The PDF parameters were derived from the results from measured indoor temperatures in social housing located in England[28].*

Figure 12 shows the magnitude and duration of positive and negative flexibility obtained when using the PDF to assign the indoor air temperature of each dwelling (due to the use of a PDF impacting the initial indoor temperature and consequently operating level of heat pumps, each run of the model may provide slightly different results).

It can be observed that:

- The magnitude of negative flexibility is smaller than shown in Figure 9 . It is due to having a 34% probability that a dwelling will be assigned an initial air temperature below the threshold of +18˚C and thus not being able to provide any demand reduction service. This is not the case for the magnitude of positive flexibility as the max temperature of the PDF is +24˚C and this is also the threshold of the model for maximum indoor air temperature.
- The duration for which the flexibility services can be provided is also impacted because all dwellings do not have the same initial indoor air temperature.

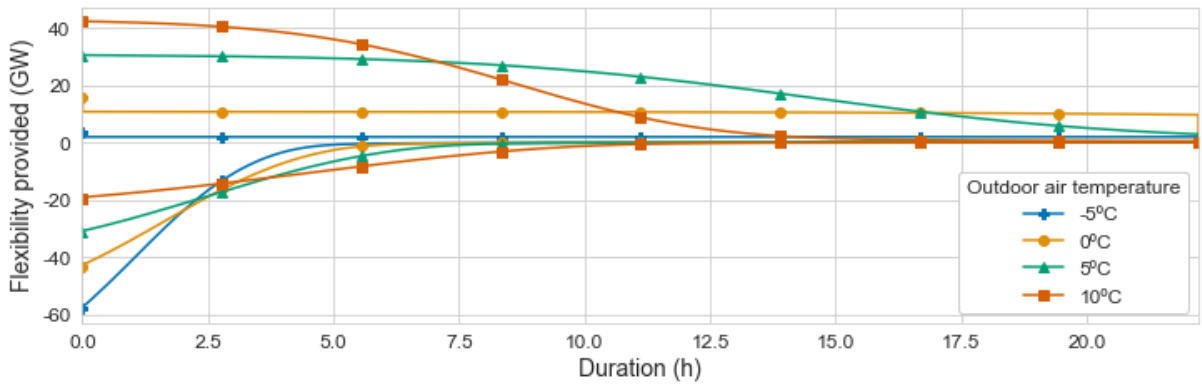

*Figure 12: Estimated magnitude and duration of flexibility services provided when the initial indoor air temperature in dwellings is based on a probability distribution function. The initial indoor air temperature of each dwelling was assigned using a PDF function which is shown in Figure 11.*

### 3.3.2 Impact of a decrease in thermal losses compared to the current configuration

In the future, energy efficiency measures are expected to be implemented and impact the thermal losses of the dwelling stock. To represent this scenario, the annual heat demand after energy efficiency measures [25] was used to model a dwelling stock with lower thermal losses. Additionally, the size of heat pumps was re-calculated for each dwelling to compensate for the heat loss at the design outdoor and indoor temperature, considering their new reduced heat loss rate.

Figure 13 compares the magnitude and duration of positive and negative flexibility services for the dwelling stock in England & Wales before and after implementing energy efficiency measures for two outdoor air temperatures of -5°C and +10°C. A decrease in the magnitude of flexibility that can be provided is observed due to a decrease in the size of the heating systems installed. Furthermore, as the thermal losses are lower, but the thermal capacity remained the same, the duration of providing flexibility is longer.

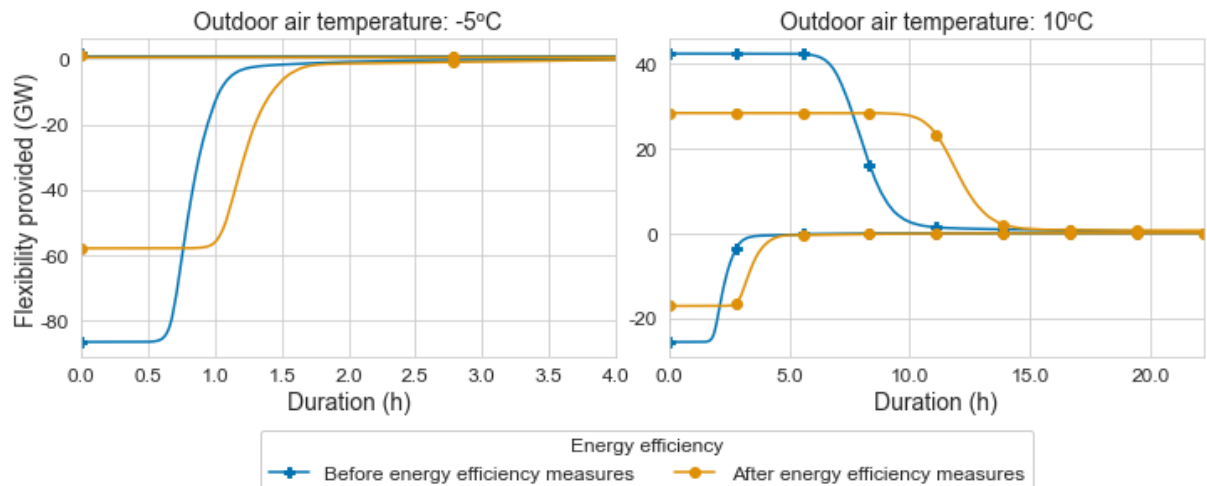

*Figure 13: Impact of energy efficiency measures on the provision of flexibility services. Comparison of the magnitude and duration of flexibility services provided for the England and Wales dwelling stock before and after implementing energy efficiency measures for two outdoor air temperatures.*

### 3.3.3 Sensitivity analysis of the thermal mass

Different types of insulation techniques could have different impacts on the thermal capacity of a building. For example, while internal insulation could reduce the usable thermal capacity of a building, external insulation could increase its usable thermal capacity. Therefore, there is uncertainty regarding the thermal capacity of future dwelling stock that will implement energy efficiency measures[29]. To investigate the impacts of such uncertainty on the available flexibility from the residential heat sector, flexibility envelop of the future housing stock were produced for three levels of thermal capacity: medium, medium + 10% and medium -10 %.

Figure 14 compares the magnitude and duration of positive and negative flexibility services for a dwelling stock which implemented energy efficiency measures for three levels of thermal capacity and two outdoor air temperatures. The thermal capacity influences the duration for which flexibility can be provided. A higher thermal capacity will result in a longer duration and vice-versa.

A 10% increase/decrease in the thermal capacity of the dwellings results in a 10% increase/decrease in the magnitude of energy that can be provided for negative flexibility at -5°C and +10°C.

A 10% increase/decrease in the thermal capacity of the dwellings results in a 9.6% increase/decrease in the magnitude of energy that can be provided for positive flexibility at +10°C.

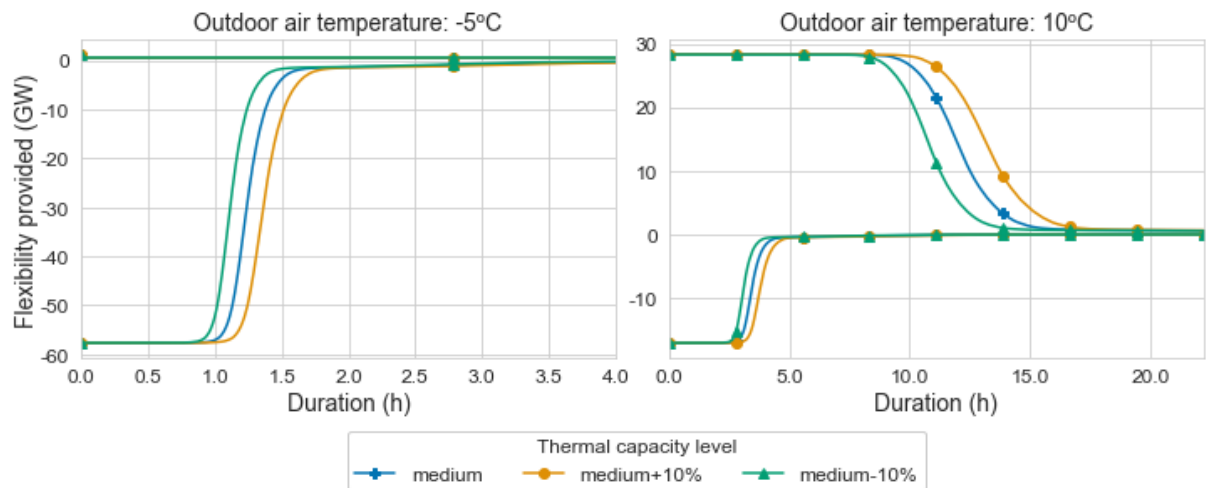

*Figure 14: Impact of thermal capacity of dwellings on the provision of flexibility services. Comparison of the magnitude and duration of flexibility services provided for the England and Wales dwelling stock with different levels of thermal capacity for two outdoor air temperatures.*

## 4 Discussion

There are several aspects of the results which can be discussed including the relation between the available flexibility and the outside air temperature, the impact of energy efficiency measures on the available flexibility and the uncertainties surrounding the approach.

The difference in the magnitude and duration of flexibility that can be provided when the outdoor air temperature and the indoor air temperature vary was highlighted in the results. As the outdoor air temperature decreases, the magnitude of positive flexibility decreases but the duration for which it can be provided increases. The opposite was observed with negative flexibility. A decrease in the initial indoor temperature of the dwellings would increase the magnitude of the positive flexibility but decrease the magnitude of the negative flexibility that could be provided. Furthermore, the flexibility services provided by the ASHPs would only be available on heating days, thus an alternative source of flexibility would be required in summer.

The potential uptakes of energy efficiency measures described in the sensitivity section (Section 3.3.3) highlighted that a decrease in the thermal losses of the buildings will lead to a lower magnitude in the flexibility services provided but an increase in their duration. However, the impact of energy efficiency measures on the thermal mass of the buildings is difficult to assess with certainty. For instance, external wall insulation may increase the thermal mass of the dwelling whereas internal wall insulation decreases it. Previous studies demonstrated the challenges of optimising thermal mass and insulation and their impacts on operational energy consumption [29,30].

Options to maintain or improve the duration of flexibility services which do not depend on insulation measures, would include installing thermal storage systems [6] and switch to district heating supply systems that could embed large thermal storage and leverage the thermal energy stored within the district heating network [31].

The approach used to quantify the flexibility from buildings have some limitations due to the uncertainties around the accuracy of the thermal parameters of the dwelling stock and the model used.

The EPC dataset used to derive the thermal losses and thermal capacitances of the residential dwelling stock is known to have limitations [32] but it currently offers the most accurate source of dwelling data in the UK. A difference of less than 10% was found between the average annual heat demand of dwelling categories calculated from EPC registers and the results from a study by the Centre for Sustainable Energy [25]. In the same paper, similar difference was found when comparing to the heat demand estimated using residential gas demand data with sub-national gas demand statistics.

A lumped parameter thermal model of building (1R 1C) was used. It is acknowledged that the accuracy of the results could be improved by utilising more complex models which consider variables such as the number of heated rooms, occupancy, schedule of appliances, solar irradiation and wind.

Further improvements of the modelling approach could be achieved by modelling additional dwelling categories and using more accurate data for the indoor air temperature of the dwellings, the thermal characteristics of dwellings, the sizing of the heat-pumps and their controls. The type of controls used could have a significant impact on the availability of the flexibility services.

# 5   Conclusions

This study aimed at quantifying the potential of the electrification of the residential heat sector and the inherent thermal energy storage of dwellings to provide demand-side flexibility services to the electric power system.

The thermal parameters of the England and Wales dwelling stock were derived from EPCs and the SAP 2012. They were used as input data to the transient model developed to quantify the flexibility from the thermal mass of dwellings. The magnitude and duration of the flexibility services available were quantified for a scenario in which 100% of the dwellings in England and Wales are equipped with ASHPs.

With a predicted uptake of residential heat pumps in the FES 2021 pathways between 28% and 80% by 2050 in the UK [1], the residential heat sector could significantly help to provide flexibility to the public electricity network. Based on these heat-pumps uptakes, our analysis showed that for England and Wales, at +5°C, between 8.5 and 24.3 GW of positive flexibility and 12 to 34 GW of negative flexibility can be provided from several minutes to few hours. As a comparison, the total DSR requirements were estimated to be between 19.2 and 44 GW and the total flexibility requirements were between 120 GW and 232 GW in the FES 2021 pathways.

At local levels, GB electricity distribution network operators (DNOs) have launched services for customers to provide DSR to help manage the constraints of the electricity distribution network. For instance, the DNO that supplies electricity to Cornwall, is currently looking for flexibility providers for the green-shaded areas in Figure 15. For November 2022, they are looking to have 678 MW available for negative flexibility at 18:00 for every day of the week. Our results showed that if all dwellings in Cornwall were using ASHPs, 457 MW of negative flexibility could be provided when the outdoor air temperature is +5°C. This represents 70% of the requirements considering our modelling assumptions.

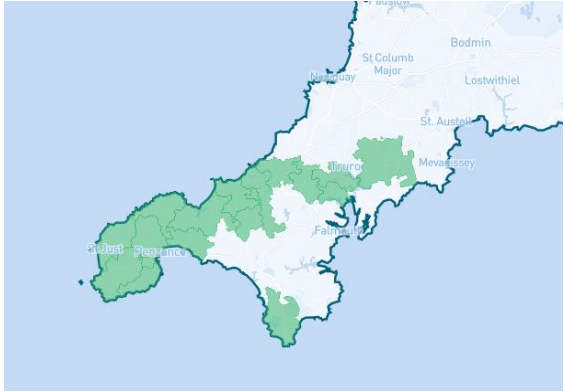

*Figure 15: Areas with flexibility requirements in Cornwall, UK. Extracted from a map showing the areas that are procuring flexibility (August 2022) source: Western Power Distribution (Distribution Network Operator of the Southwest of England).*

There are current market and technical barriers for households to access the flexibility market. To enable and increase the provision of flexibility services from the residential heat sector, a number of measures and changes in the electricity market will need to take place [33]. Furthermore, the infrastructure to tap into this resource will need to be installed including remote control of heating systems and data metering solutions.

# 6   Code availability

The source code of this study is available on Github at:
https://github.com/AlexandreLab/flexibilitydwellings.

# 7   Data availability

The results of this study are available to download on the UKERC Energy Data Centre website at https://ukerc.rl.ac.uk/DC/cgi-bin/edc_search.pl?GoButton=Detail&WantComp=282&&RELATED=1

# 8   Acknowledgements

This work was supported by the UK Engineering and Physical Sciences Research Council (EPSRC) through UKERC (EP/S029575/1) and MISSION project (EP/S001492/1).